\documentstyle[epsf,rotate]{mn}

\newcommand{\ie}{{\sl i.e. }}

\newcommand{\Msun}{\mbox{$M_{\odot}$}}
\newcommand{\Rsun}{\mbox{R$_{\odot}$}}

\newcommand{\rchis}{\mbox{$\chi^{2}_{\nu}~$}}
\newcommand{\kms}{$\,$km$\,$s$^{-1}$}

\title[The mass of X-Ray Nova Scorpii 1994 (=GRO~J1655--40)]
{The mass of X-Ray Nova Scorpii 1994 (=GRO~J1655--40)}
\author[T.~Shahbaz, F.~van der Hooft, J.~Casares, P.A.~Charles and
J.~van Paradijs]{
T.~Shahbaz$^{1}$, F.~van der Hooft$^{2}$, J.~Casares$^{3}$,
P.A.~Charles$^{1}$ and J.~van Paradijs$^{2}$ \\
$^{1}$University of Oxford, Department of Astrophysics, Nuclear Physics 
Building, Keble Road, Oxford, OX1 3RH, England \\
$^{2}$Astronomical Institute ``Anton Pannekoek'', University of Amsterdam
and Center for High Energy Astrophysics, Kruislaan, 403, 1098 SJ \\
Amsterdam, The Netherlands \\
$^{3}$Instituto de Astrof\'\i{}sica de Canarias 38200 La Laguna, 
Tenerife, Spain}

\begin{document}

\maketitle

\begin{abstract}

\noindent
We have obtained high and intermediate resolution optical spectra of the
black-hole candidate Nova Sco 1994 in May/June 1998, when the source was
in complete (X-ray) quiescence. We measure the radial velocity curve of
the secondary star and obtain a semi-amplitude of
$K_{2}$=215.5$\pm$2.4\kms, which is 6 per cent lower than the only
previously determined value. This new value for $K_{2}$ thus reduces the
binary mass function to $f(M)$ = 2.73$\pm$0.09 \Msun. Using only the high
resolution spectra we constrain the rotational broadening of the
secondary star, $v\sin~i$, to lie in the range 82.9--94.9 \kms (95 per
cent confidence) and thus constrain the binary mass ratio to lie in the 
range 0.337--0.436 (95 per cent confidence).
We can also combine our results with published limits for
the binary inclination to constrain the mass of the compact object and
secondary star to the ranges 5.5 -- 7.9  and 1.7 -- 3.3
\Msun respectively (95 per cent confidence). Finally, we report on
the detection of the Lithium resonance line at 6707.8\AA, with an
equivalent width of 55$\pm$8 m\AA.

\end{abstract}


\section{Introduction}

The soft X-ray transient, Nova Sco 1994 (=GRO~J1655--40) has been studied
extensively over the past 3 years in X-rays and at optical wavelengths
[Tingay et al. 1995, Bailyn et al. 1995a, Harmon et al., 1995, Hjellming
\& Rupen 1995, Bailyn et al. 1995b, Paciesas et al. 1996, van der Hooft
et al., 1997, van der Hooft et al., 1998 (hereafter vdH98)]. Strong
evidence that the compact object in GRO~J1655-40 is a black hole was
presented by Bailyn et al.\ (1995b) who initially established a
spectroscopic period of $2.601\pm0.027$ days, and determined a mass
function $f(M)=3.16\pm0.15$
\Msun. An improved value of $f(M)=3.24\pm0.09$ \Msun\ was presented by
Orosz \& Bailyn (1997; hereafter OB), who measured the radial velocity
semi-amplitude of the secondary star to be $K_{2}=228.2\pm2.2$ \kms
and classified it as an F2--F6$\sc iv$ type star.
Both vdH98 and OB analyzed their quiescent optical light curves
of Nova Sco 1994 which they combined with the OB value for $K_{2}$ 
to show that the 
black hole has a mass ($M_{1}$) in the range 6.29--7.60 \Msun.

However, in calculating the radial velocity semi-amplitude, OB used both
quiescent data (taken in 1996 February 24--25) \emph{and} outburst data
(taken in 1995 April 30--May 4). Using outburst data in this way can lead to
an incorrect result as substantial X-ray heating of the secondary star
shifts the `effective centre' of mass of the star (see Wade \&
Horne 1988 and Phillips, Shahbaz \& Podsiadlowski 1998). This results in
a significant distortion of the radial velocity curve and hence
a spuriously high velocity semi-amplitude. The masses of
the binary components derived from this will therefore be incorrect.

In this paper we determine the radial velocity curve of the secondary
star in Nova Sco 1994 using only data taken when the system was in
quiescence. From our high resolution spectroscopy we determine the
rotational broadening of the secondary star, which, when combined with
the new value for the mass function and limits to the binary inclination
(vdH98). allows us to determine more precisely the
individual component masses.

\begin{table}
\caption{NTT log of Nova Sco 1994 observations}
\begin{center}
\begin{tabular}{lccc}
Date      &  Exp. time       & Resolution  & Wavelength range \\
          &                  &             &                 \\ 
29/5/1998 & 12$\times$1800s  & 0.83~\AA\   & 6131--6764~\AA\ \\
30/5/1998 & 2$\times$1800s   & 4.2~\AA\    & 4452--7015~\AA\ \\
31/5/1998 & 2$\times$1800s   & 4.2~\AA\    & 4452--7015~\AA\ \\
 2/6/1998 & 13$\times$1800s  & 0.83~\AA\   & 6131--6764~\AA\
\end{tabular}
\end{center}
\end{table}	

\section{Observations and Data Reduction}

Intermediate and high resolution optical spectra of Nova Sco 1994 were
obtained on 1998 May 28, 29, 30 and June 1 with the 3.5-m New Technology
Telescope (NTT) at the European Southern Observatory (ESO) in Chile using
the ESO Multi Mode Instrument (EMMI). We used the red arm with an
order-separating OG 530 filter and gratings \#6 and \#8 which gave
dispersions of 0.31~\AA\ per pixel and 1.26~\AA\ per pixel respectively
(see Table 1). The TEK 2048$\times$2048 CCD was used, binned with a
factor two in the spatial direction in order to reduce the readout noise.
The dispersion direction was not binned. Very good seeing allowed us to use 
a slit width of $0\farcs8$ 
which resulted in spectral resolutions of 0.83~\AA and 4.2~\AA\ for
the gratings \#6 and \#8 respectively (see Table 1). Cu-Ar arc spectra
were taken for wavelength calibration. Template field stars of a variety
of spectral types were also observed with intrinsic rotational broadening
much less than the resolution of our high resolution data. 

The data reduction and analysis was performed using the Starlink {\sc
figaro} package, the {\sc pamela} routines of K.\,Horne and the {\sc
molly} package of T.\,R.\ Marsh. Removal of the individual bias signal
was achieved through subtraction of the mean overscan level on each
frame. Small scale pixel-to-pixel sensitivity variations were removed
with a flat-field frame prepared from observations of a tungsten lamp.
One-dimensional spectra were extracted using the optimal-extraction
algorithm of Horne (1986), and calibration of the wavelength scale was
achieved using 5th order polynomial fits which gave an rms scatter of
0.02~\AA. The stability of the final calibration was verified with the OH
sky line at 6562.8\AA\ whose position was accurate to within 0.1 \AA.

\begin{table}
\caption{Sine-wave fits to the radial velocity curvs}
\begin{center}
\begin{tabular}{lll}
Parameter  		        & NTT (this paper) & NTT + CTIO (OB) \\
				&                  &          \\
Orbital period$^{*}$ (days)     &  2.62168         &  2.62168 \\
T$_{0}^{*}$ (HJD 2,440,000+)    & 9838.4198        & 9838.4198 \\
$K_{2}$ velocity (km~s$^{-1}$)  & 215.5 $\pm$ 2.4  & 233.3 $\pm$ 4.9 \\
$\gamma$ velocity (km~s$^{-1}$) & $-$141.9 $\pm$ 1.3 & $-$142.4 $\pm$ 2.9 \\
phase shift                     & 0.05 $\pm$ 0.003 & 0.03 $\pm$ 0.005 \\
$\chi^{2}_{\nu}$		& 0.99             & 4.9 \\
$f(M)$ ($M_{\odot}$)            & 2.73 $\pm$ 0.09  & 3.46 $\pm$ 0.22
\end{tabular}
\end{center}
\noindent
$^{*}$Taken from van der Hooft et al. (1998) and fixed.\\
\noindent
1-$\sigma$ uncertainties are given, see text.
\end{table}

\section{The radial velocity of the companion star}

Our first task was to measure the radial velocity of the F-type secondary
star. The radial velocities were measured from the spectra by the method
of cross-correlation (Tonry \& Davis 1979) with a template star. Prior to
cross-correlation the spectra were interpolated onto a logarithmic
wavelength scale (pixel size 14.5 \kms) using a $\sin\,x/x$ interpolation
scheme to minimize data smoothing (Stover et al. 1980), and then
normalised. We masked the interstellar feature at 6280~\AA\ and H$\alpha$
before correlating the spectra. We used HR2927 as our template star after
we had broadened it by 90 \kms in order to simulate the spectrum of Nova
Sco 1994 (a more accurate value for the rotational broadening is derived
in the next section).
The radial velocity of the template star (derived using the position of
the H$\alpha$ absorption line to be $-$6.7 \kms) was then added to the
radial velocities of Nova Sco 1994.

Using the orbital ephemeris given by vdH98 we phase-folded the
heliocentric radial velocities and then fitted a sine wave (see Table 2
and Figure 1), obtaining a $\chi^{2}_{\nu}$ of 0.99. We then combined the
data with the quiescent radial velocity data of OB and repeated the
fitting procedure obtaining a significantly worse value of
$\chi^{2}_{\nu}$=4.9. Note that our high resolution data has a factor of
4 better spectral resolution than that of OB; the uncertainty in our
individual radial velocity measurements being typically 6\kms.

\begin{table}
\caption{Results for the optimal subtraction of the companion star}
\begin{center}
\begin{tabular}{lccc}
Name   & Sp. Type    & $\chi^{2}_{\nu}$ &  $f$ \\ \\
HR3325 & F4$\sc iii$ &  2.5 &  0.99 \\
HR2927 & F6$\sc iii$ &  1.7 &  0.98 \\   
HR870  & F7$\sc iv $ &  2.2 &  0.99 \\
HR9046 & F8$\sc iv $ &  3.1 &  0.94 \\
HR9057 & F8$\sc iii$ &  2.3 &  0.96 \\
HR6192 & G3$\sc iii$ &  2.6 &  0.82 \\
HR7864 & G5$\sc iii$ &  3.2 &  0.88 \\
HR7281 & G8$\sc iii$ &  4.4 &  0.62 \\
\end{tabular}
\end{center}
\end{table}	

\section{The spectral type and rotational broadening of the companion star}

We determine the spectral type of the companion star by minimizing the
residuals after subtracting different template star spectra from the
Doppler-corrected average spectrum. This method is sensitive to the
rotational broadening $v\sin~i$ and the fractional contribution of the
companion star to the total flux. The template stars we use are in the
spectral range F0--F8, $\sc v$--$\sc iii$ and were obtained during this
observing run but also from previous runs at La Palma and 
with comparable dispersion.

First we determined the velocity shift of the individual high resolution
spectra of Nova Sco 1994 with respect to each template star spectrum, 
then Doppler-averaged to the rest frame of the template star (see
section 3). In order to determine the rotational broadening, $v\sin~i$,
we follow the procedure described by Marsh, Robinson \& Wood (1994).
Basically we subtracted a constant, $f$, representing the fraction of
light from the template star, multiplied by a rotationally broadened
version of that template star.
The optimal values of $v\sin~i$ and $f$ are obtained by minimising
$\chi^{2}$ (see Table 3). The above analysis was performed in the
spectral ranges 6380--6520 \AA\, and 6600--6750 \AA\, which excludes 
H$\alpha$. This was the only region common to all the
templates stars and the high resolution spectra of Nova Sco 1994. We used a
linear limb-darkening coefficient of 0.52 (Al-Naimiy 1978)
appropriate for 6500 \AA\ and an effective temperature of 6500 K (typical
for a F star). Figure 3 shows the results for those templates with $f < 1$.
It should be noted that the main sequence template stars all gave $f > 1$,
as also found by OB. From Table 3 it can be seen that the 
minimum $\chi^{2}$ occurs at spectral type F6 with a $v\sin~i$ of
91.2$\pm$3.7 \kms. The 90 per cent confidence levels shown were obtained
using a Monte Carlo simulation (see section 5 for details), after the
\rchis of the fit was multiplied by a scaling factor so that the 
$\chi^{2}_{\nu,\rm min}$ was 1 (Lampton et al. 1976).

\begin{table}
\caption{The rotational broadening of the companion star and the binary 
mass ratio. 90 per cent confidence levels are given.}
\begin{center}
\begin{tabular}{lcc}
Limb darkening   &    $v\sin~i$      &  $q$(=$M_{2}/M_{1}$)   \\
coefficient ($u$) &   (km~s$^{-1}$)  &                        \\
                 &                   &                        \\
0.00   	         & 86.5 ($+$3.3, $-$3.6) &  0.360 ($+$0.028, $-$0.033) \\
0.52             & 91.2 ($+$3.7, $-$3.7) &  0.400 ($+$0.036, $-$0.030) 
\end{tabular}
\end{center}
\end{table}	

The analysis above assumes that the limb-darkening coefficient
appropriate for the radiation in the line is the same as for the
continuum. However, in reality this is not the case; the absorption lines
in early-type stars will have core limb-darkening coefficients much less
than that appropriate for the continuum (Collins \& Truax 1995).
In order to determine the extreme limits for $v\sin~i$ we also repeated
the above analysis for the F6 template star using zero limb-darkening and
found $v\sin~i$=86$^{+3.3}_{-3.6}$ \kms (90 per cent confidence). We
thus constrain $v\sin~i$ to lie in the range 82.9--94.9 \kms (95 per
cent confidence; since we have two one-tailed distributions representing
the upper and lower limits).

As can be seen from Figure 3 (top) the determination of the spectral type
depends critically on the choice of template spectral type and luminosity
class. Even if we had many template stars we would find that using
stars of the nominally same spectral class and type would give differing
results. In Figure 3 we also show a fit to the data; this is
what would be expected if there were no systematic effects.

In order to estimate this systematic effect in our analysis of Nova Sco
1994 we performed a Monte Carlo simulation involving the same analysis
but now using a a template star of known spectral type (F5$\sc iv$;
HR870) as the target. To this target we added noise to produce a spectrum
of comparable quality to our Doppler-averaged Nova Sco 1994 spectrum. We
then repeated the broadening and optimal subtraction procedure using the
same 21 templates star as was used for the Nova Sco 1994 analysis,
thereby determining the best fit. We found that we could determine the
spectral type of the trial template star to within 1 sub-class. Figure 3
(bottom) shows a single simulation; note the systematic effects, \ie the
scatter within each spectral type bin.

\section{The binary system parameters}

Since the companion star fills its Roche lobe and is synchronised with the
binary motion, the rotational broadening provides a direct measurement of the
binary mass ratio, $q$ (=$M_{2}$/$M_{1}$), through the expression 
$v\sin~i = [K_{2} (1 + q ) 0.49q^{2/3}] / [0.6q^{2/3} + \ln(1+q)^{1/3}$]
(Horne, Wade \& Szkody 1986).
Substituting our values for $K_2$ and $v\sin~i$ we calculate $q$ (see
Table 4). In order to determine the uncertainty in $q$ we used a Monte
Carlo simulation, in which we draw random values for the observed
quantities which follow a given distribution, with mean and variance the
same as the observed values. For $K_2$ the random distribution to taken
to be Gaussian as the uncertainty is symmetric about the mean value.
However for $v\sin~i$ the uncertainty is asymmetric and so we had to
determine the actual distribution numerically. This was done by first
calculating the maximum likelihood distribution using the actual
$\chi^{2}$ fit values (after rescaling the
\rchis values so that the 
$\chi^{2}_{\nu,\rm min}$ was 1) and then
determining the cumulative probability distribution. By picking random
values (from a uniform distribution) for the probability we obtain random
values for $v\sin~i$. Given the uncertainty in the limb darkening
coefficient and hence $v\sin~i$, we find $q$ to lie in the range 0.337 --
0.436 (95 per cent confidence).

Using our values for $K_{2}$ and $q$ with the orbital period, $P$, and
the binary inclination, $i$, we can determine the masses of the compact
object, $M_{1}$ and the companion, $M_{2}$ star using $PK_{2}^{3}/2\pi G
= M_{1}\sin^{3}i/(1+q)^{2} $. The binary inclination is given by vdH98,
and lies in the range 63.7$^\circ$ and 70.7$^\circ$. It should be noted
that the solutions for the binary inclination and hence binary masses,
determined by vdH98 depend on the flaring angle of the accretion disc.
Here we use the range for the binary inclination which 
encompasses the range of
flaring disc angles assumed by vdH98. Table 5 shows the values we obtain
for the masses which depend on the limb-darkening coefficient used to
determine $v\sin~i$ and the binary inclination. The 90 per cent
confidence level was determined using a Monte Carlo simulation (see
Figure 4) similar to that described above. Given the uncertainties in the
limb-darkening coefficient (and thus in $v\sin~i$ and $q$) and the binary
inclination, we find $M_{1}$ and $M_{2}$ to lie in the range 5.5 -- 7.9
\Msun and 1.7 -- 3.3 \Msun respectively (95 per cent confidence).

By modelling the optical light curves of Nova Sco 1994 vdH98 and OB have
also determined $q$. They find $q$ to lie in the range 0.24--0.42
(3-$\sigma$) and 0.289--0.385 (3-$\sigma$) respectively. (However, it
should be noted that the errors given by OB are only internal statistical
errors, whereas the analysis of vdH98 includes the
systematic uncertainties.) Using optical spectroscopy (see section 5), we
find $q$ to lie in the range 0.337--0.436 (95 per cent confidence). Also
from optical spectroscopy, the projected radial velocity semi-amplitude
for the compact object has been determined by Soria et al., (1998) to be
76.2$\pm$7.5 \kms. We can use this along with our value for $K_{2}$ (see
section 3) to determine $q$; $q$=0.354$\pm$0.035 (1-$\sigma$). As one can
see, the four independent measurements for $q$ above are fully consistent
with each other.
This consistency is also seen in the determination of the
binary inclination; van der Hooft 1997, vdH98, and OB determine 
values for $i$ which are all also consistent with each other.

\begin{table}
\caption{The masses of the binary components. 90 per cent confidence levels 
are given.}
\begin{center}
\begin{tabular}{lccc}
Limb darkening & Inclination &    $M_{1}$    &    $M_{2}$    \\
coefficient (u) &             & ($M_{\odot}$) & ($M_{\odot}$) \\
               &             &               &               \\
0.00   	       & 63.7  	     & 6.4--7.4      & 2.0--2.8      \\
0.52   	       & 63.7  	     & 6.8--7.9      & 2.4--3.3      \\
               &             &               &               \\
0.00   	       & 70.7  	     & 5.5--6.3      & 1.7--2.4      \\
0.52   	       & 70.7  	     & 5.8--6.7      & 2.1--2.9      
\end{tabular}
\end{center}
\end{table}

\section{The detection of lithium?}

High lithium abundances appear to be a common feature of late-type
secondary stars in neutron stars and black hole SXTs, a result with is
totally unexpected given the post-main sequence dilution and the mass
transfer history of these stars (Mart\'\i{}n et al., 1994). In order to
counteract the depletion of lithium, caused by convective mixing, it is
necessary to invoke a process to produce lithium. Several mechanisms have
been put forward to explain the high lithium abundance in the neutron
star and black hole SXTs; lithium synthesis in the supernova explosion
that formed the compact object (Woosley et al., 1990), or $\alpha-\alpha$
reactions during the strong SXT outbursts (Mart\'\i{}n et al., 1994).

We can use the ratio of Li~$\sc i/$Ca~$\sc i$ equivalent width (=R$_{\rm
Li/Ca}$), which is not sensitive to veiling for deriving upper limits to
the surface Li abundance (see Figure 5). For the other SXTs A0620-00,
V404 Cyg and Cen X--4 R$_{\bf Li/Ca}$ has been found to be $>$ 1
(Mart\'\i{}n et al., 1994); for Nova Sco 1994 we find R$_{\bf
Li/Ca}$=0.48$\pm$0.08. The equivalent widths of Li~$\sc i$ and Ca~$\sc i$
are 55$\pm$8 m\AA\ and 114$\pm$10 m\AA\ rrespectively.

The interpretation of the lithium abundance in Nova Sco 1994 is somewhat
difficult. The secondary stars in SXTs are tidally locked and therefore
have high rotational velocities, hence, convective depletion may be
inhibited by reducing the angular momentum transport in the base of the
convection zone. One might expect there to be a correlation between
rotation rate and lithium abundance (Pallavicini, Cerruti-Sola \& Duncan
1987), since convective mixing may cause additional depletion if lithium
is taken to layers where it can be burnt. However, the link between
rotation and lithium in subgiant stars is not yet well established; de
Medeiros, do Nascimento Jr \& Mayor (1997) have found that the lithium
abundance is independent of rotational velocity.
Indeed the sample tabulated by Balachandran (1990) shows a remarkable
range of lithium equivalent width (2--100 m\AA) within F6$\sc iv$ stars
but there is {\it no} correlation with rotational velocity. Thus it is
difficult to compare the actual lithium equivalent width in Nova Sco 1994
with that in ``normal'' F6$\sc iv$ stars, such as our template star
HR2927 in Fig. 5. All we can say is that we have detected lithium in Nova
Sco 1994, but whether it is intrinsic to the secondary star remains
unclear. Spallation appears to be the most obvious mechanism as it is the
most probable cause of the high Li abundance seen in other SXT, but in
Nova Sco 1994 mechanisms related to the spectral type, luminosity of the
star and magnetic braking have to be considered as well.

\section{Discussion}

From Figure 2 it can be seen that after removal of the secondary star
from the average spectrum of Nova Sco 1994, there is {\it narrow}
residual $H\alpha$ emission (EW=10.0$\pm$0.5\AA; FWHM=146 \kms) which
could be either due to chromospheric activity on the secondary star,
and/or X-ray heating on the inner face or emission from the bright spot.
H$\alpha$ emission arising from the quiescent accretion discs in SXTs
usually have a FWHM of a few thousand \kms; the residual feature we see
has a FWHM of only a few hundred \kms (e.g. Casares et al., 1997), so it
must arise from regions on the secondary star or the outer edge of the
accretion disc (Marsh et al,, 1994). Our data do not have sufficient
phase resolution to allow us to distinguish between the above
possibilities; only a detailed phase-resolved radial velocity study of
the narrow component will solve its origin. 

However, it is interesting to note that the narrow emission can be 
powered solely by X-ray heating.
An upper limit to the X-ray flux on Nova Sco 1994 is set by $RXTE$ ASM
observations at $\leq F_{X}=3.1\times 10^{-10}$ erg~cm$^{2}$~s$^{-1}$.
The X-ray irradiation at the secondary star would then be $\leq 2.3\times
10^{10}$ erg~cm$^{2}$~s$^{-1}$, where we have used a distance of 3.2 kpc
(Hjellming \& Rupen 1995) and a binary separation of 16.6 \Rsun. At such
levels the secondary would be receiving sufficient energy to power the
narrow H$\alpha$ emission.

In calculating the radial velocity semi-amplitude of the secondary
star OB used both quiescent (1996 February 24--25)
\emph{and} outburst data (1995 April 30--May 4), 
and fitted the resulting
curve with a sine wave. It has been known, especially in studies of dwarf
novae (Wade \& Horne 1988 and Davey \& Smith 1992) that using outburst
data in this way leads to incorrect results. The effect of substantial
heating of the secondary causes the centre of light, given by the
strength of the absorption lines, to shift from the centre of mass. This
results in a significant distortion of the radial velocity curve leading
to a spuriously high radial velocity semi-amplitude (Davey \& Smith
1992). The binary masses derived from this will
therefore be incorrect.

Phillips, Shahbaz \& Podsiadlowski (1998) have estimated $K_{2}$ by
fitting only the outburst radial velocity data of OB. They use a crude
model based on X-ray heating of the secondary star and estimate $K_{2}$
to lie in the range 194--214 \kms (90 per cent confidence), assuming
$q$=0.33. Note that this range compares well with our value of
$K_{2}$=215.5$\pm$2.4 \kms obtained by fitting our high resolution radial
velocity data of Nova Sco 1994 in complete quiescence. We therefore
conclude that the masses of the Nova Sco 1994 system are somewhat lower
than the OB value, but the compact object mass is still sufficiently high
($>3$\Msun) to imply that it is a black hole system.

\section*{Acknowledgements}

We would like to thank Jerry Orosz for providing the OB radial velocity
data and Sam Phillips and Mark Seaborne for useful discussions.

\begin{figure*}
\rotate[l]{\epsfxsize=500pt \epsfbox[00 00 700 750]{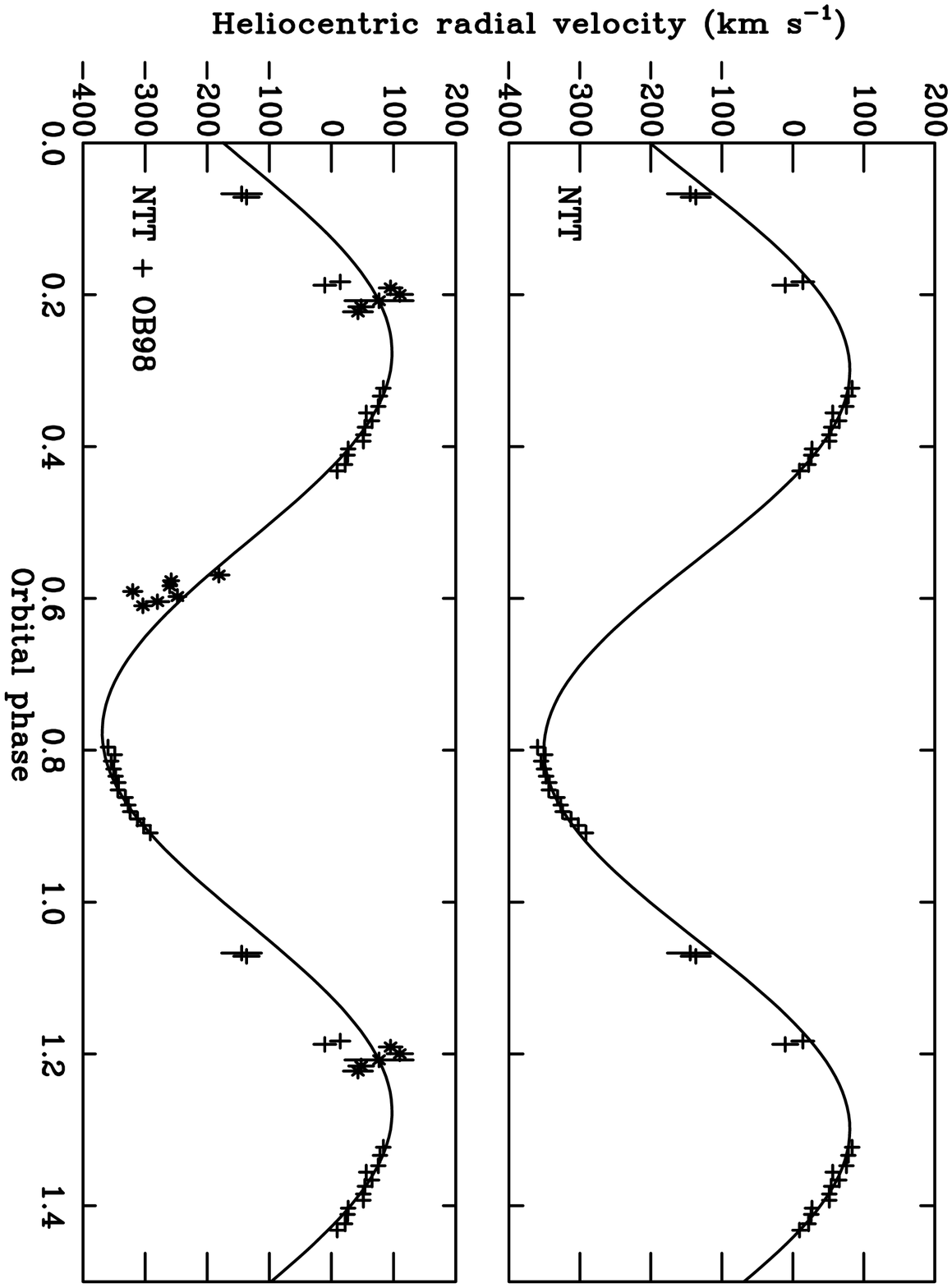}}
\caption{Top panel: The radial velocity curve of the secondary star obtain
using only our NTT data. The solid line shows a sinusoidal fit to the
data. Bottom panel: The radial velocity curve of the secondary star
obtain using our NTT data (crosses) and also the data of OB (stars). 
The solid lines
in each panel show a sinusoidal fit to the data. The data have been
folded on the orbital ephemeris given by van der Hooft et al., (1998),
and 1.5 cycles orbital are shown.}
\end{figure*}

\begin{figure*}
\rotate[l]{\epsfxsize=500pt \epsfbox[00 00 700 750]{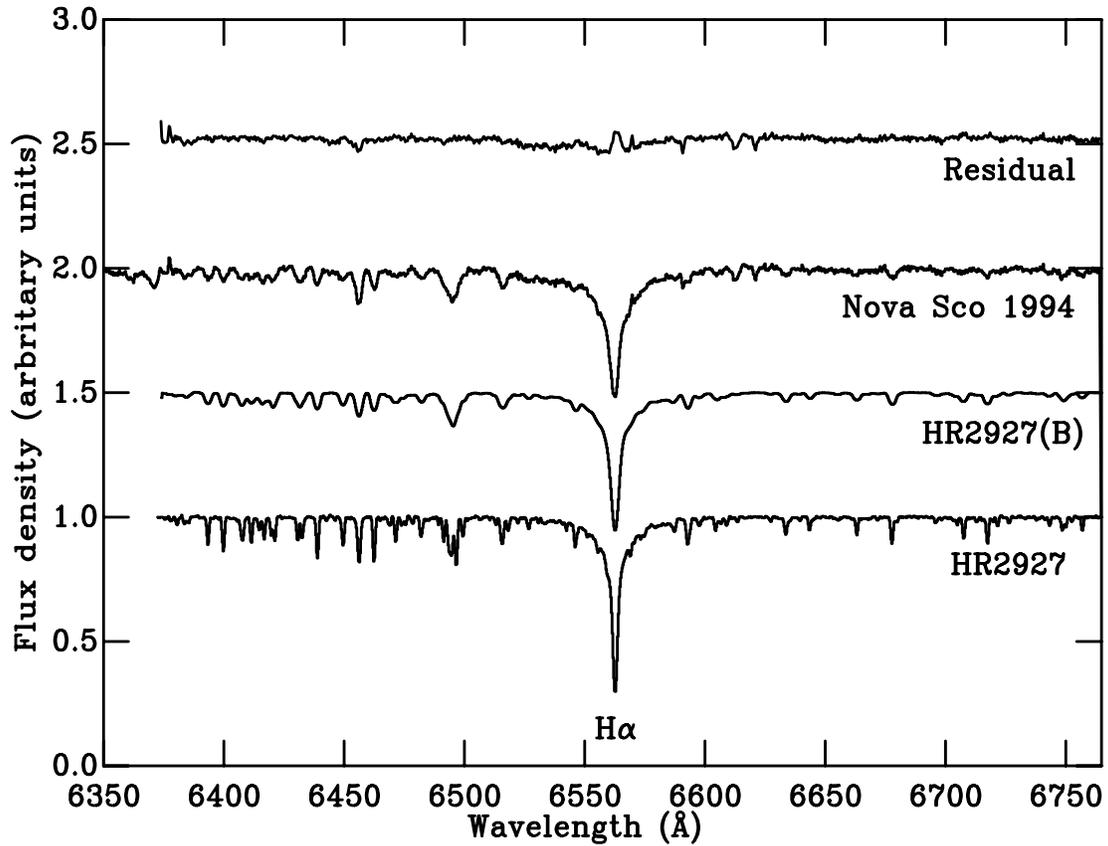}}
\caption{The results of the optimal subtraction. From top to bottom:
the residual spectrum of Nova Sco 1994 after subtracting the template
star times $f$=0.98, the variance-weighted Doppler averaged spectrum of
Nova Sco 1994, the template F6$\sc iv$ star (HR2927) broadened by
91.2\kms, and the template F6$\sc iv$ star again to show the
narrow absorption lines.  The spectra have been normalized and shifted
vertically for clarity.}
\end{figure*}

\begin{figure*}
\rotate[l]{\epsfxsize=500pt \epsfbox[00 00 700 750]{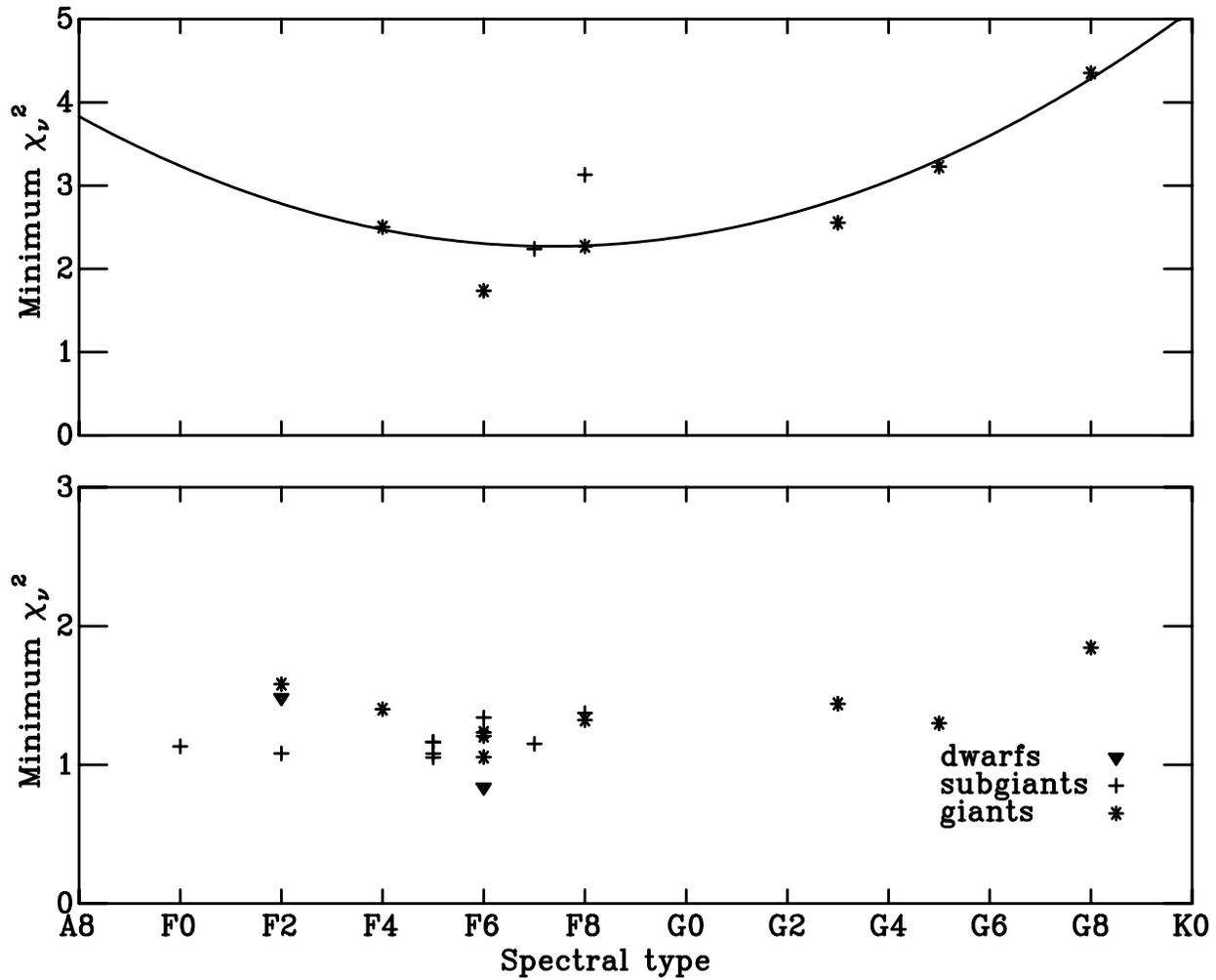}}
\caption{Top: The $\chi^{2}_{\nu}$ of the optimal subtraction analysis as a
function of the spectral type of the template star used. The solid line
is a quadratic fit to the data. Bottom: Same as above but for a sample
star of known spectral type (F5). Different symbols indicate different
luminosity classes: triangles for dwarfs as shown. 
Only template stars that gave $f<1$ in the analysis are shown (see text). }
\end{figure*}

\begin{figure*}
\rotate[l]{\epsfxsize=500pt \epsfbox[00 00 700 750]{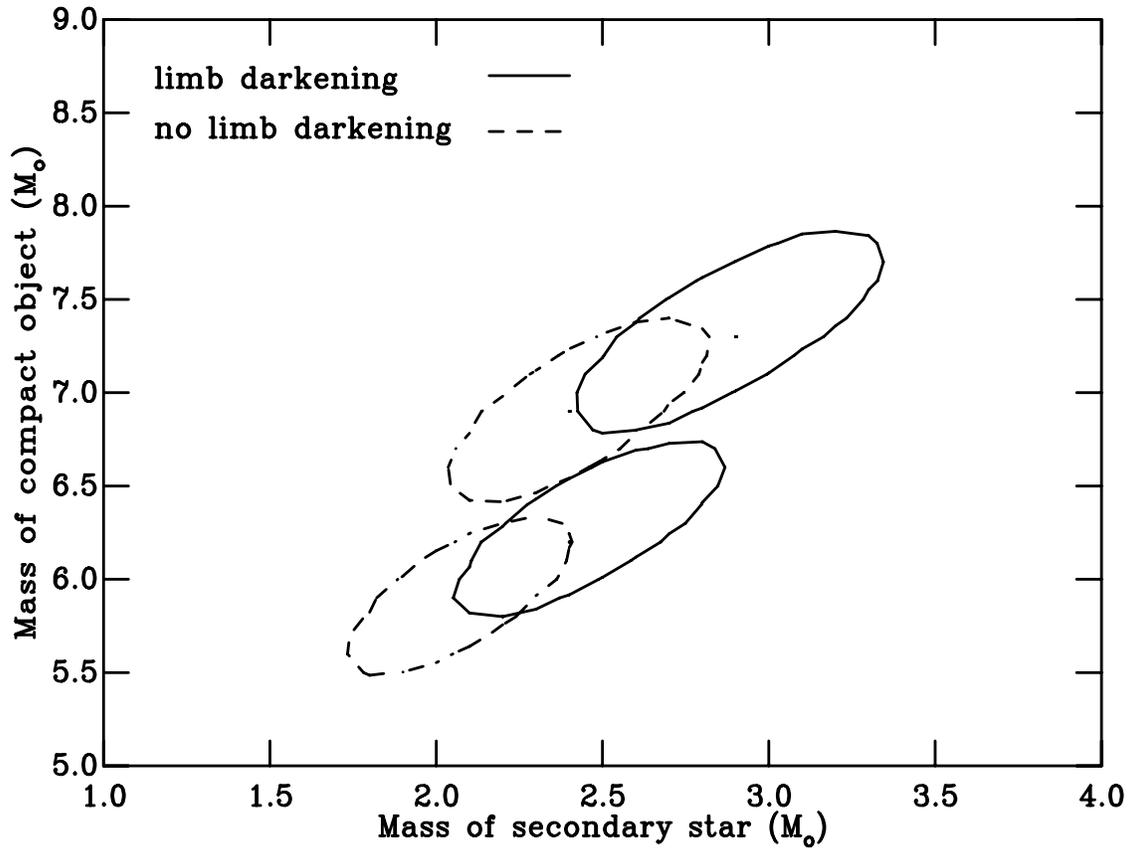}}
\caption{The mass of the binary components obtained using a Monte Carlo
simulation (see section 4). 90 per cent confidence regions are shown.
The solid ellipses are regions obtained using the continuum value for the 
limb-darkening coefficient, where as the dashed ellipses are regions
obtained using no limb darkening. The top and bottom 
two ellipses are for a binary inclinations of 63.7$^{\circ}$ and 
70.7$^{\circ}$ (van der Hooft et al., 1998). }
\end{figure*}

\begin{figure*}
\rotate[l]{\epsfxsize=500pt \epsfbox[00 00 700 750]{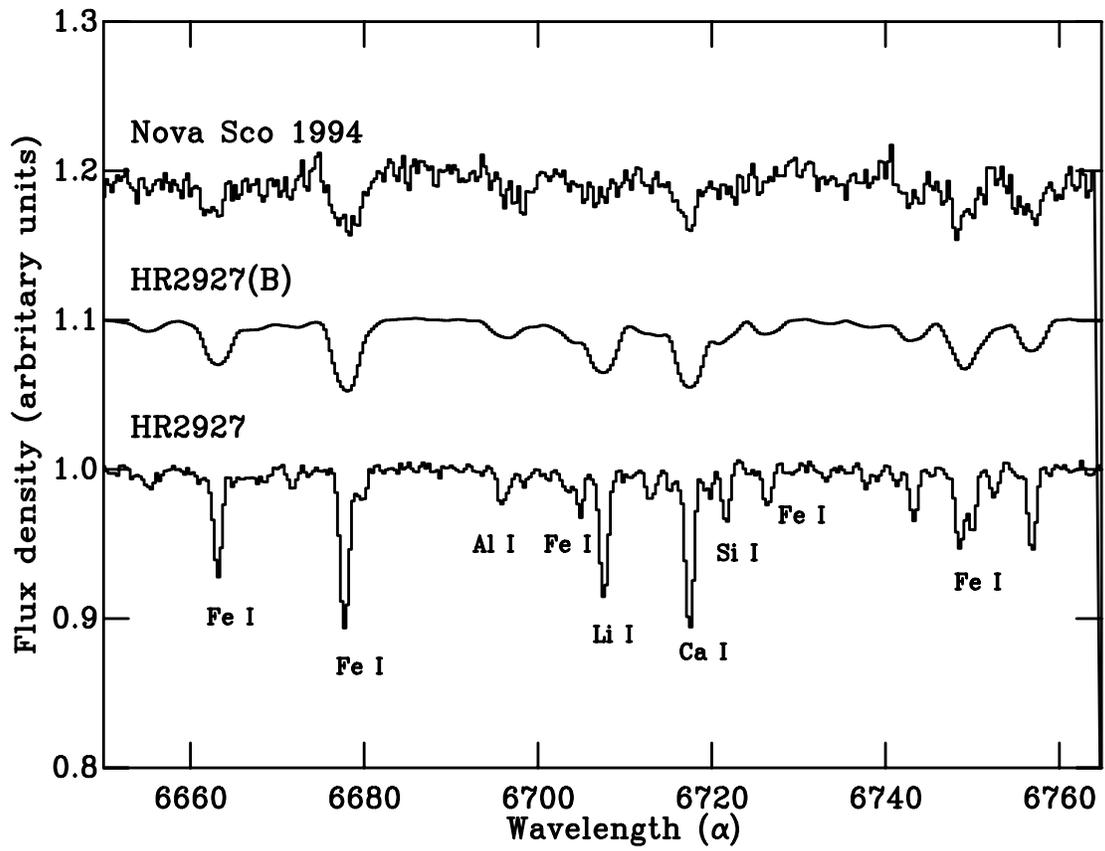}}
\caption{From top to bottom: The Doppler averaged spectrum of Nova Sco
1994, a template star (F6$\sc iv$) broadening by 91 \kms to match Nova
Sco 1994, the same template, to show positions of the absorption lines
more clearly. The spectra have been offset for clarity.}
\end{figure*}

\end{document}